%% file: paper.tex
\input mnd.tex
\input psfig


\def\simless{\mathbin{\lower 3pt\hbox
   {$\rlap{\raise 5pt\hbox{$\char'074$}}\mathchar"7218$}}}   
\def\simgreat{\mathbin{\lower 3pt\hbox
   {$\rlap{\raise 5pt\hbox{$\char'076$}}\mathchar"7218$}}}   
\def \AA    #1 {A\&A, #1, }
\def \AJ    #1 {AJ, #1, }
\def \ApJ   #1 {ApJ, #1, }
\def \MNRAS #1 {MNRAS, #1, }
\def \Nat   #1 {Nature, #1, }

\def\etal{{\rm et al.}}

\def\solmas{{M$_\odot$}}
\def\solm{{M_\odot}}
\def\solrad{{R$_\odot$}}

\def\rst {$R_{\star}$}
\def\mst {$M_{\star}$}
\def\vipc {$V_{\rm ipc}$}
\def\rmag {$R_{\rm hole}$}
\def\rs {R_{\star}}
\def\ms {M_{\star}}
\def\vip {V_{\rm ipc}}
\def\mip {{\rm M}_{\rm ipc}}
\def\mipc {${\rm M}_{\rm ipc}$}

\def\mtr {{\rm M}_{\rm tracks}}
\def\mtra {${\rm M}_{\rm tracks}$}
\def\rmg {R_{\rm hole}}
\def\vstr {$V_{\rm stream}$}
\def\vst {V_{\rm stream}}

\pageoffset{-2.5pc}{0pc}
\Autonumber


\pubyear{1997}

\begintopmatter
%

\title{Magnetospheric accretion and PMS stellar masses}
\author{Ian A. Bonnell$^1$, Kester W. Smith$^2$, Michael R. Meyer$^3$, Christopher A. Tout$^1$, Daniel F. M. Folha$^{2,4}$ and James P. Emerson$^2$}
\affiliation{$^1$ Institute of Astronomy, Madingley Road,
Cambridge CB3 0HA}
\affiliation{$^2$ Queen Mary and Westfield College, Mile End Road, London,
E1 4NS}
\affiliation{$^3$ Hubble Fellow, Steward Observatory, University of Tucson, AZ 85721, USA }
\affiliation{$^4$ Centro de Astrofisica da Universidade do Porto, Rua do Campo
Alegre, 823, 4150 Porto, Portugal}

\shortauthor{I. A. Bonnell et. al.}
\shorttitle{PMS stellar masses}

\acceptedline{Accepted by MNRAS}

\abstract
We present a method of determining lower limits on the masses of
pre-main-sequence (PMS) stars and so constraining the PMS evolutionary
tracks. This method uses the red-shifted absorption feature 
observed in some emission-line profiles of T Tauri stars 
indicative of infall.  The maximum velocity of the
accreting material measures the potential energy at the stellar
surface, which, combined with an observational determination of the
stellar radius, yields the stellar mass. This estimate is a lower limit
owing to uncertainties in the geometry and projection effects.
Using available data, we show that the computed lower limits can be
larger than the masses derived from PMS evolutionary tracks for $M
\simless 0.5 \solm$.  Our analysis also supports the notion 
that accretion streams do not impact near the stellar poles but 
probably hit the stellar surface at moderate latitudes.

\keywords stars: accretion, accretion discs -- stars: formation -- stars: luminosity function, mass function.

\maketitle 

\section{Introduction} 

\tx A complete understanding of the star formation process
requires the ability to predict how the properties of young stars
depend on the initial conditions of star formation.  Despite
significant observational effort in the last two decades, a key
experimental problem remains. That is reliable determination of the
masses and ages of pre-main sequence (PMS) stars.  Stellar masses are
needed to investigate how the initial mass function (IMF) varies from
region to region as a function of initial conditions (i.e. cloud
parameters).  With reliable stellar ages in hand, we can begin to
address quantitatively important questions of early stellar evolution.
Presently, a common method used to determine masses and ages of young
stars is to place them in an H-R diagram and compare their positions
with theoretical evolutionary tracks (eg Hillenbrand~1997).
Considerable theoretical work has been done recently that has advanced
our understanding of PMS evolution (D'Antona \& Mazzitelli~1994
[DM94]; Swenson \etal~1994).  Yet there remain significant differences
in the tracks owing to alternative treatments of convection, opacities
and the zero-point of the calculated ages. An additional complication
is on-going accretion during the PMS, which can alter the path of a
star during its early evolution (Hartman, Cassen \& Kenyon~1997; Seis,
Forestini \& Bertout~1997).  Without adequate observational
constraints, it is nearly impossible to determine what the correct
treatment should be.

The discrepancies in the PMS tracks are largest for the lowest-mass
stars (Hartigan, Strom, \& Strom~1994).  
It is precisely this end of the stellar mass distribution which is
the most uncertain both amongst field stars as well as in star-forming 
regions (eg Kroupa, Tout, \& Gilmore~1993; Luhmann \& Rieke~1998).  
Bona fide brown dwarf objects have recently been discovered as companions
to field stars (Nakajima et al. 1995) and as free--floating members of 
young clusters (Rebolo, Zapatero-Osorio \& Martin ~1995). However the 
frequency of such objects remains unknown.  
Young clusters found in regions of star-formation 
provide one of the best opportunities to determine 
the relative contribution of low-mass stars and brown dwarfs to 
the cluster IMF because substellar objects cool as they age, 
becoming extremely difficult to detect when old. 
Uncovering local maxima or minima in the IMF near the stellar/substellar 
boundary would point to the existance of characteristic masses in the 
formation process (Adams \& Fatuzzo~1996). Because of the importance of the 
low-mass end of the IMF in star-formation theory, 
it is imperative to establish reliable PMS
evolutionary tracks and thus PMS masses.  To do this, we need independent 
determinations of PMS masses to constrain the tracks.

The best techniques available for directly determining stellar masses
involve the study of binary star systems. 
Eclipsing binary systems provide the best estimates of stellar
masses as well as radii (Popper~1980).  Unfortunately, only a few such 
systems are known that are sufficiently young to provide a 
test of the PMS
evolutionary tracks (eg Casey \etal~1998) and even these do not probe the
lowest masses.  Astrometric determination of orbital parameters for 
visual PMS binary star systems is another promising technique 
(Ghez \etal~1995), but useful constraints are still several years away.  
Dynamical mass ratios can be derived for double-lined spectroscopic 
binaries and compared with theoretical mass ratios in order to test
the PMS tracks (Lee~1992).  However most of the known double-lined PMS 
SBs have primary stars with masses greater than solar and mass ratios 
near one.  We therefore need other constraints,
even if they can provide only lower or upper limits on the stellar
mass. Such limits can be obtained from the dynamical information
available from the material accreting on to T~Tauri and other young
stars.

Accretion on to young stars occurs through a viscous circumstellar
disc, where angular momentum transport outwards permits the inward
mass transport (Pringle~1981; Lin \& Pringle~1990).  There is a
growing body of observational evidence indicating that the accretion
of material on to young stars occurs via magnetospheric accretion
columns which extend several stellar radii from the PMS photosphere to
the inner edge of a circumstellar disc (Edwards~1997). Material at
this point falls freely along the field lines on to the stellar
surface. Evidence for this picture comes predominantly from the
observed redshifted absorption component (several hundred
km$\,$s$^{-1}$) seen in the line profiles of the higher Balmer,
Pa$\beta$, Br$\gamma$, HeI and NaD emission lines (Walker~1972,
Appenzeller, Reitermann \& Stahl~1988; Edwards \etal~1994; Hartmann,
Hewett \& Calvet~1994; Folha, Emerson \& Calvet~1997, Folha \&
Emerson~1998).  Models of the IR colours of T Tauri stars imply that
the circumstellar discs have inner holes comparable (but slightly
interior) to the magnetospheric radius which is typically estimated at
6-8 R$_*$ (Kenyon, Yi \& Hartmann~1996; Meyer, Calvet \&
Hillenbrand~1997; Armitage, Clarke \& Tout~1998). The magnetospheric
accretion paradigm also provides an attractive explanation of the
relatively slow spin rates of T Tauri stars (Bouvier \etal~1993;
Edwards \etal~1993).  The stellar magnetic field interacts with parts
of the disc that are spinning slower than the star (at radii beyond
corotation) and the resulting magnetic braking of those stars with
discs (the actively accreting CTTS) can explain why they have longer
periods than those (the non-accreting WTTS) stars without
circumstellar discs (Cameron \& Campbell~1993; Armitage \&
Clarke~1996).

In this paper, we investigate how the observed red-shifted absorption
components, known as inverse P-Cygni (IPC) profiles, can be used as a
measurement of the depth of the potential well of the star and hence,
combined with an observational determination of the stellar radius,
yield an estimate of a lower limit on the stellar mass. Section~2
describes the method while Section~3 provides examples, using
available data, of the mass estimates from this method and compares
them with published estimates based on the PMS tracks. Section~4 shows
how the lower limits on the stellar mass could in principle constrain
the accretion stream geometry. A discussion and summary is presented
in \S~5.

\section{Constraining PMS masses}

\tx The IPC profiles in the emission line spectra of classical T~Tauri (CTTS)
stars offer a direct measurement of the depth of the potential well of
the star. The large infall velocities seen in the absorption profiles
are generally assumed to arise from material in free-fall from 
several stellar radii.  This near radial infall is believed to be
caused by the disruption of the accretion disc by the stellar magnetic
field  (Konigl~1991, Edwards \etal~1994). The 
velocity of the stream, as it impacts the stellar surface then
directly measures the difference in potential energy from the radius
at which the disc is disrupted to the stellar surface:
$$ {1\over 2} \vst^2 = {{G \ms \over \rs} - {G \ms \over \rmg}},
\eqno(1)$$ where \vstr\ is the velocity of the accretion stream at impact, 
\mst\ is the stellar mass, \rst\ is the stellar radius, \rmag\ is the 
magnetospheric radius at which the disc is disrupted and $G$ is the
gravitational constant.  
Knowing these quantities, we can
evaluate the stellar mass as
$$\ms ={\vst^2 \rs \over 2G} \biggl(1-{\rs\over\rmg}\biggr)^{-1}.\eqno(2)$$
The rotational energy of the disc matter at
\rmag\ is assumed to be dissipated as 
the disc matter couples on to the star's magnetic field lines, and is therefore
neglected in equation~(1). This energy 
is at most half the potential energy at \rmag\ and thus its contribution
to equation~(1) is equivalent to the disc being disrupted at $2\ \rmg$. Mass
estimates incorporating the rotational energy will therefore lie in between the
two limits we calculate below; i) that the matter falls in from infinity; 
or ii) from \rmag.

Although the value of $\rs/\rmg$ is uncertain, it can be estimated by
comparing the infrared excess emission expected from a circumstellar
disc with and without an inner hole. Kenyon, Yi \& Hartmann~(1996)
estimate that $\rmg/\rs \approx 4$ (with acceptable values between 3
and 5).   Similar results were obtained by Meyer \etal~(1997) who 
estimate that $2 < \rmg/\rs < 6$ for 
a sample of T Tauri stars located in the Taurus dark cloud.  In
practice, because we are looking for a lower limit on the stellar
mass, and as $\rs/\rmg$ is relatively small, we calculate the stellar
mass using equation~(2) assuming either that the material falls in from
infinity or from five stellar radii ($\rmg \approx 5 \rs$).  We note
that the case of infall from infinity provides a strict lower limit to
the mass resulting in values $0.8$ times those  obtained in the 
$5 \rs$ calculation.  

We see from equation~(2) that we can make an estimate of the stellar
mass from just the stellar radius and the impact velocity of the
accretion stream.  The stellar radius can be calculated directly from
the stellar luminosity and $T_{\rm eff}$. It is thus independent of
the PMS tracks, but does depend on 
observers' ability to correct for reddening, establish the
spectral type (i.e. photospheric temperature), 
and separate stellar from accretion luminosity. 
The velocity itself can be estimated from the red-shifted 
absorption profile.  There is an added
complication that the accretion stream is not necessarily parallel to 
our line of sight so that there is a projection effect in the velocity.
The observed IPC velocity relates to the impact velocity as
$$\vip = \vst \times {\rm cos} \theta, \eqno(3)$$ where $\theta$
measures the angle between our line of sight and the direction of the
stream on impact.  This implies that the observed \vipc\ will be a
lower limit on the true impact velocity and again gives us a lower limit
on the stellar mass.  Furthermore, as the accretion streams most
probably rotate with the star, they will be obscured by the star
over part of the rotation period (eg Smith \etal~1997) and the observed
\vipc\ will be variable (Edwards \etal~1994).

In the above we have not considered the exact mechanism by which the disc
matter couples to the magnetic field and the accompanying torques. This
simplification neglects any azimuthal motions of the infalling matter
as it follows the field lines. Fortunately, any such deviations
also ensure that the mass estimates are lower limits. In fact, 
all of the unknowns involved in measuring the star's gravitational
potential well, the distance from where the material falls, the 
projection of the stream along our line of sight, that the
stream impacts radially, and even the assumption
that the stream's maximum velocity is on the stellar surface and not
at some distance from the star, make our estimate a lower limit.
Therefore we should, in general, underestimate the gravitational
potential well of a star. Given  an accurate determination
of the stellar radius from its luminosity and temperature, the
mass determined from this method is a lower limit on the
true stellar mass.
  
\section{Comparison of IPC mass limits and masses from PMS tracks}

\table{1}{D}{{\bf Table 1.} \rm Comparison of T Tauri masses from PMS tracks 
and lower limits from IPC profiles. The radii (in \solrad), and PMS track 
masses (in \solmas) are taken from Hartigan \etal~1995 using the tracks 
from D'Antona \& Mazzitelli~1994. The IPC profile velocities (in km s$^{-1}$) 
are from $^1$Edwards \etal~1994, from $^2$Folha \& Emerson~(1998) and 
from $^3$Appenzeller \etal~1988. Two values of \vipc\ are used:  these are
the deepest absorption, \vipc$({\rm deep}$) corresponding to \mipc,  and the  the maximum velocity 
in the absorption profile \vipc$({\rm max})$ corresponding to \mipc$^*$ (see text). The masses, 
$\mip/\solm$,
are calculated on the assumptions that the matter falls in from infinity,
\mipc$(\infty)$, and from $5 \rs$, \mipc$(5\rs)$. The 
last three columns list the rotational periods in days, $v{\rm sin} i$
in km s$^{-1}$, and the inclination angles (from Bouvier
\etal~1995), adapted to the stellar radii from Hartigan \etal~1995). 
An inclination angles of 90$\deg$ is assigned to GM Aur where the above
comparison yields ${\rm sin}i > 1$} {$$\centerline {\vbox {\tt
\halign {\strut ##\hfil &
\hfil ## \hfil & \hfil ## \hfil & \hfil ## \hfil & \hfil ## \hfil &
\hfil ## \hfil & \hfil ## \hfil & \hfil ## \hfil & \hfil ## \hfil &
\hfil ## \hfil & \hfil ## \hfil & \hfil ## \hfil & \hfil ## \hfil \cr
\noalign{\hrule} 
{\rm Star~~} & {\rm Radius} & {\rm M$_{\rm tracks}$} & {\rm line} 
 & {\vipc$({\rm deep}$)} & {\mipc$({\infty})$} & {\mipc$(5\rs)$} 
 & {\vipc$({\rm max})$} & {\mipc$(\infty)^*$} & {\mipc$(5\rs)^*$} & P$_{\rm rot}$ & $v{\rm sin} i$ & {$i$} \cr
 \noalign{\hrule} 
{\rm AA Tau} & {\rm 1.8} & {\rm 0.38} &H${\beta}$& {\rm 220$^1$} & {\rm 0.23} 
   & {\rm 0.28} & {\rm 280$^1$} & {\rm 0.34} & {\rm 0.41} & {\rm 8.2} 
	& {\rm 11.4} & {\rm 81}\cr 
{\rm BP Tau} & {\rm 1.9} & {\rm 0.45} &H${\delta}$& {\rm 290$^1$} & {\rm 0.42}
   & {\rm 0.51} & {\rm 400$^1$} & {\rm 0.8} & {\rm 0.96} & {\rm 7.6} 
	& {\rm 7.8} & {\rm 38}\cr
{\rm CW Tau} & {\rm 2.2} & {\rm 1.03} &Br${\gamma}$& {\rm 170$^2$} 
   & {\rm 0.17} & {\rm 0.20} & {\rm 260$^2$} & {\rm 0.40} & {\rm 0.48} \cr 
{\rm DK Tau} & {\rm 2.7} & {\rm 0.38} &H${\gamma}$& {\rm 280$^1$}  
   & {\rm 0.56}	& {\rm 0.67} & {\rm 315$^1$} & {\rm 0.70} & {\rm 0.84} 
   & {\rm 8.4} & {\rm 11.4} & {\rm 43}\cr 
{\rm DL Tau} & {\rm 1.9} & {\rm 0.37} &{\rm Na D}& {\rm 240$^1$} & {\rm 0.29}
   & {\rm 0.34} & {\rm 315$^1$} & {\rm 0.50} & {\rm 0.60} & \cr 
{\rm DN Tau} & {\rm 2.2} & {\rm 0.42} &H${\beta}$& {\rm 160$^1$} 
   & {\rm 0.15} & {\rm 0.18} & {\rm 205$^1$} & {\rm 0.25} & {\rm 0.30} 
   & {\rm 6.0} & {\rm 8.1} & {\rm 27}\cr 
{\rm DO Tau} & {\rm 2.4} & {\rm 0.31} &Pa${\beta}$& {\rm 175$^2$} 
   & {\rm 0.19} & {\rm 0.23} & {\rm 235$^2$} &{\rm 0.34} & {\rm 0.41} \cr
{\rm DR Tau} & {\rm 2.7} & {\rm 0.38} &{\rm HeI}& {\rm 225$^1$} & {\rm 0.36} 
   & {\rm 0.43} &{\rm 315$^1$} & {\rm 0.71} & {\rm 0.85} & {} & {} & {} \cr 
{} & {} & {} &H${\delta}$& {\rm 330$^3$} & {\rm 0.77} & {\rm 0.93} & {} & {} & {}\cr 
{\rm DS Tau} & {\rm 1.6} & {\rm 1.28} &H${\delta}$& {\rm 265$^1$} & {\rm 0.29}
   & {\rm 0.35} & {\rm 370$^1$} & {\rm 0.56} & {\rm 0.68} &\cr 
{\rm FM Tau} & {\rm 1.6} & {\rm 0.15} &Pa${\beta}$& {\rm 120$^{2}$} 
   & {\rm 0.06} & {\rm 0.07} & {\rm 210$^2$} & {\rm 0.18} & {\rm 0.21} &\cr  
{\rm GI Tau} & {\rm 2.5} & {\rm 0.30} &Pa${\beta}$& {\rm 225$^{2}$} 
   & {\rm 0.33} & {\rm 0.40} & {\rm 350$^2$} & {\rm 0.81} & {\rm 0.97} 
   & {\rm 7.2} & {\rm 11.2} & {\rm 40}\cr
{\rm GK Tau} & {\rm 2.2} & {\rm 0.41} &H${\beta}$& {\rm 250$^1$} & {\rm 0.36}
   & {\rm 0.43} & {\rm 280$^1$} & {\rm 0.46} & {\rm 0.55} & {\rm 4.65} 
	& {\rm 18.7} &{\rm 52}\cr 
{\rm GM Aur} & {\rm 1.6} & {\rm 0.52} &Pa${\beta}$& {\rm 250$^2$} & {\rm 0.26}
   & {\rm 0.31} & {\rm 315$^2$} & {\rm 0.41} & {\rm 0.49} & {\rm 12.0} 
	& {\rm 12.4} & {\rm 90}\cr
\noalign{\hrule}
}}}$$}

\tx In order to ascertain how useful this method is for
constraining PMS masses, we compare masses derived from PMS tracks (\mtra)
with the lower limits on the stellar masses calculated with equation~(2) (\mipc).
Table~1 shows this comparison for 13 CTTS stars chosen from the stars
studied by Hartigan \etal~(1995).  This sample includes all stars 
with observed IPC profiles (from Edwards \etal~1994, 
Folha \& Emerson~1998 and in one case Appenzeller \etal~1988) 
that have also been surveyed and found  {\it not} to
have any companions within separations $0.1-2.0 ''$ (Mathieu~1994
and references therein;
Simon \& Prato~1995).  
Companions in this separation range are
unresolved in the Hartigan \etal~(1995) study and hence would introduce
errors into the determination of the stellar luminosity and thus the
radii and PMS track masses (Ghez, White \& Simon~1997). This highlights 
another advantage of the technique; it can be applied to {\it single} 
stars and not just binary systems in contrast to the methods for 
determining directly stellar masses discussed above.   The radii and
masses (from D'Antona \& Mazzitelli~1994 tracks adopting Alexander 
opacities and CM convection)
are taken from Hartigan \etal~1995.  These authors have attempted to
deredden the stars and remove the effects of accretion
luminosity before calculating the stellar parameters.

\vbox{
\figure{1}{D}{0mm}{\vskip-0.5truein\centerline{\vbox{\psfig{figure=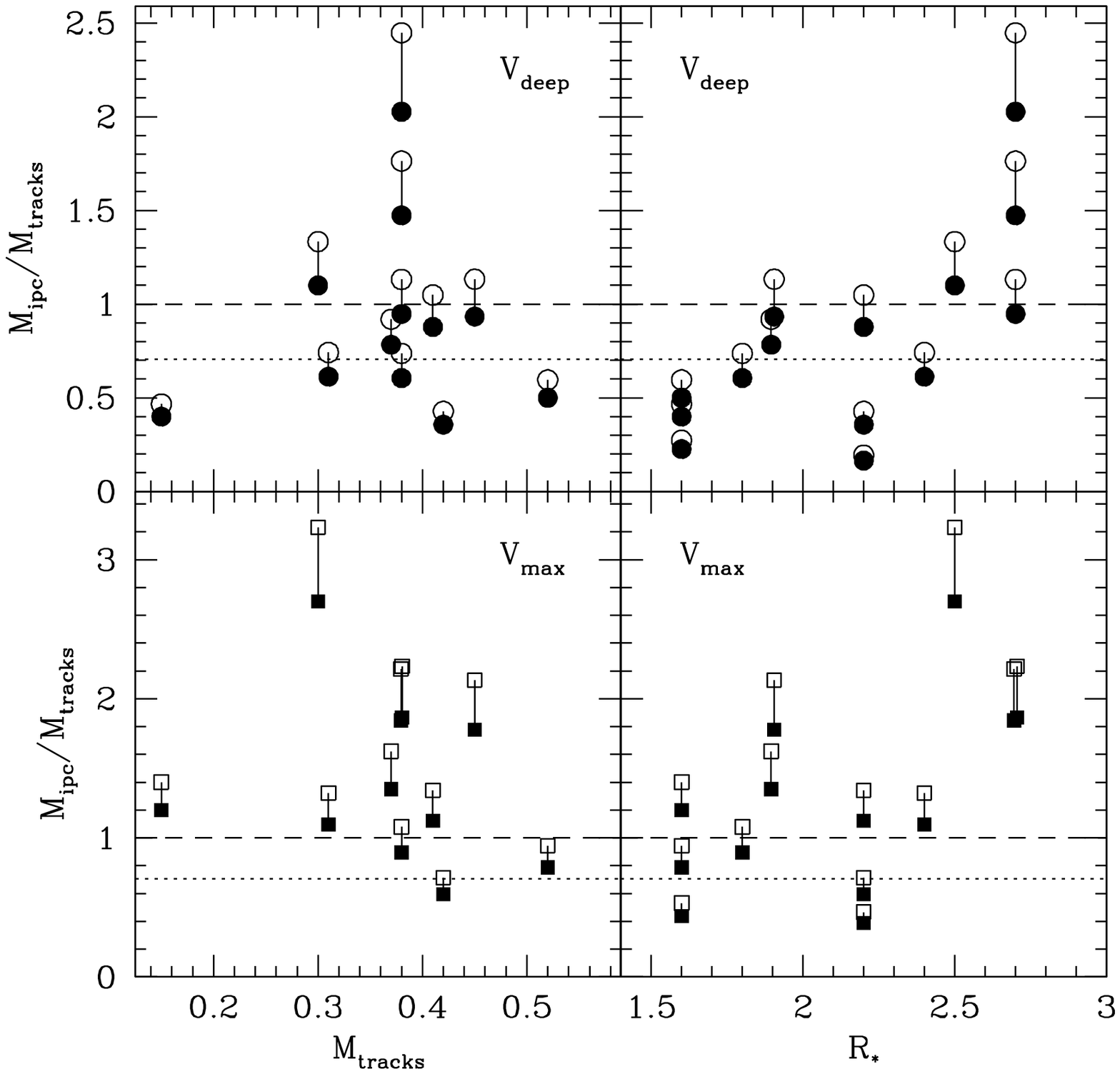,width=5.75truein,height=5.75truein,rwidth=5.5truein,rheight=5.5truein}}} 
\break\noindent
{\bf Figure 1.} The ratio of the dynamical mass estimate, \mipc, to
the mass derived from the PMS tracks, \mtra, is plotted against $\mtr/\solm$
(left panels) and against the stellar radii in \solrad\
(right panels). The top panels (circles) use the deepest part of the
absorption profile as being the characteristic \vipc, while the bottom
panels (squares)assume that \vipc\ is the maximum velocity in the
absorption profile. The filled symbols assume the matter falls in from
infinity while the open symbols assume that the mass falls in from
five stellar radii. As \mipc\ is a strict lower limit, the ratio of
$\mip /\mtr$ should always be less than 1 (dashed line) if the PMS
track masses are correct. The dotted line represents the expected
ratio for a projection of 45 degrees.}}

The velocities \vipc\ are estimates based on the IPC profiles seen in
the Balmer, NaD and HeI lines (Edwards \etal~1994; Appenzeller
\etal~1988), and in the Paschen$\beta$ and Brackett$\gamma$ lines
(Folha \& Emerson~1998). The use of different lines to measure
the impact velocities is non-ideal as it may introduce uncertainties due
to radiative transfer effects, but is necessary because of the paucity of 
available IPC profiles. For each star, two estimates of the velocity
were made, the first from the deepest part of the
absorption feature and the second from the maximum velocity of the
absorption (taken to be where the absorption profile meets the
continuum).  
These estimates are most probably lower limits on the
velocity of the infalling material when it impacts the
stellar surface.  Modelling of the IPC profile with radiative transfer
(Muzerolle, Calvet \& Hartmann~1998) shows that the impact
velocity typically corresponds to the red-most extreme part of the
absorption component.  The minimum in the IPC profiles in Muzerolle
\etal~(1998) generally corresponds to a
\vipc\ that is 10 to 20 per cent lower than the characteristic
velocity \vstr.  We prefer to use both estimates of \vipc\ (quoted in
Table~1) as limits of the true characteristic velocity to avoid any
uncertainties in the details of the line formation. For
each star, the maximum velocity observed (of each type) was used.  In
general, IPC profiles are variable (eg Edwards \etal~1994; Smith
\etal~1997) with higher velocity IPC profiles presumably arising when
the accretion stream is most closely aligned to our line of
sight. These velocities are therefore lower limits on the actual
impact velocity of the accretion stream.  In this context it is worth
noting the two different estimates of the IPC velocity profile of DR
Tau (from Edwards \etal~1994 and Appenzeller \etal~1988) in Table~1.   
The IPC profile can even disappear completely when the stream is
behind the star. This implies that the magnetospheric accretion
geometry is complex and is not an axisymmetric ring aligned with the
rotation axis.  The dynamical mass estimates (lower limits) based on
the IPC velocity measurements are given in columns~5 and~6 of Table~1
assuming that the stream velocity, \vstr, is given by the position of
the minimum of the absorption feature, and that the matter falls in
from infinity ($\rs / \rmg = 0$) or from five stellar radii ($ \rs /
\rmg = 0.2$). The corresponding cases for the maximum velocity seen in
the IPC profile are given in columns~8 and~9. The photometric
(rotation periods, $v\ {\rm sin} i$, and inclinations of the stars,
derived by Bouvier~1995, are given in column~10.

It is difficult to quantify the uncertainties in our mass
determinations presented in Table~1. First we consider the errors in
the stellar radii derived from the observational determination of the
stellar luminosity and effective temperature.  Propagating errors in
the photometry, reddening, distance modulus, bolometric corrections
(i.e. luminosity), and spectral types (i.e. photospheric
temperatures), we estimate that the radii are accurate to within
20--30 per cent (in agreement with the analysis of Kenyon \&
Hartmann~1995).  We can estimate the systematic uncertainties in the
radii by comparing the derived values from different observational
determinations of the stellar parameters (Hartigan \etal~1994;
Gullbring \etal~1998).  Both studies rely on independent spectra and
use different corrections for the photometry. The stellar radii thus
derived independently agree remarkably well, with differences at the
10--15 per cent level.

Uncertainties in our estimates of \vipc\ are approximately 10 per
cent.  Furthermore, we neglect radiative transfer effects and the
errors associated with estimating the maximum absorption velocity in a
noisy spectra. Both of these should tend to reduce the observed
velocity from the true impact velocity. Thus, as we are deriving lower
limits, uncertainties in our estimates of \vipc\ from the data should
be smaller than the difference between our two limits.  Combining the
errors in the radii with the 10 per cent error in estimating the
velocities, this corresponds to a maximum uncertainty in our lower
limits of $\pm 45$ per cent.

Of course the errors in deriving the stellar luminosity and
temperature also result in uncertainties in the mass estimates from
the PMS tracks.  Because the mass tracks are generally vertical during
the Hyashi contraction phase for late--type stars, errors in spectral
type and conversion to effective temperature dominate the errors in
estimating stellar masses from PMS evolutionary models.  Typical
errors of $\pm 1$ spectral subclass translate into errors of $\pm
0.02$ dex in log(T$_{\rm eff}$), resulting in relative errors of $\pm 0.1
M_{\odot}$ in stellar mass estimates for young ($< 10$ Myr) late-type
(K--M) stars for the DM94 tracks (Hillenbrand~1997).  Apart from
problems with the tracks themselves (which we attempt to probe with
this technique), different approaches for placing stars in the H--R
diagram taken by different groups could be an additional source of
systematic error.  The masses reported by Gullbring et al.~(1998)
(derived from the same DM94 tracks) generally agree with those
reported by Hartigan et al.~(1995) except in those cases where a
significantly different spectral type was adopted.  Gullbring et
al.~1997 adopted a later spectral type for DS Tau than HEG (K5 vs. K2)
and an earlier spectral type for GI Tau (K6 vs. M0).

The ratio of the dynamical mass estimate, \mipc, to the track mass,
\mtra, is plotted in Figure~1 against the PMS track mass and the
radius. The ratio is plotted for both mass estimates based on the two
characteristic velocities and on the hole size in the disc (matter
falling in from infinity (filled) and from 5 stellar radii (open)). As
the dynamical mass estimates are lower limits on the stellar masses,
the ratio plotted should be less than one for all stars. We see that
this is not the case for some of the stars when we use the lower
characteristic velocity (upper panels) and with most stars if we use
the maximum velocity seen in the IPC profile (lower panels). The three
stars for which this discrepancy is most striking (DK Tau, DR Tau, and
GI Tau) are all unusual and have larger radii than the other
stars. They should be treated with some caution. DK Tau is a
relatively wide binary ($2.5''$ separation) which should, in
principle, be uncontaminated by it's companion. Contributions from a
companion can lead to a larger luminosity (radius) and hence mass
using the IPC method while not significantly effecting the mass
derived from the tracks.  GI Tau has $\mip < \mtr$ according to the
stellar parameters of Gullbring
\etal~(1998) while the high veiling of DR Tau makes it's stellar
parameters somewhat uncertain.  The other stars commonly have mass
limits that are similar to or lower than the track mass when the
lower estimates for \vipc\ are used. 
If we use the maximum velocity seen in absorption, then almost 
all the derived dynamical mass limits have $\mip \simgreat \mtr$.

Furthermore, some of the stars should be seen with
relatively large projection angles, so we would expect
them to have $\mip \simless 0.7 \mtr$ (for a projection of 45 degrees).  
Thus, from the fact that most of the stars with masses $\ms \simless 0.5 \solm$
have $\mip \simgreat 0.7 \mtr$, even when the velocity is estimated
from the deepest part of the absorption, we deduce that the
DM94 tracks potentially underestimate the true stellar masses.
This problem is aggravated if the disc is truncated at smaller radii,
resulting in higher estimates of \mipc.
Therefore, we can conclude that the dynamical
mass estimates provide a useful lower limit on the stellar mass
and thus a constraint on the PMS evolutionary tracks.

Other PMS tracks
give different results. For example, the tracks of Swenson
\etal~(1994) give systematically higher masses (see Hillenbrand~1997 
for comparison) which would compare more favourably with our upper-limits. 
However this technique can in principle only reject a set of tracks
for which the predicted masses were too low.  As long as the track 
masses are comfortably above the lower limits set by \mipc\ we cannot
prefer one set of tracks over another. Nonetheless, a 
direct comparison of different PMS track
masses (including the effects of accretion) with mass limits derived
using the method described here for a statistically significant sample
of PMS stars will assist in discarding unviable PMS evolutionary tracks.

\section{Inclination and accretion stream geometry}

\vbox{
\figure{2}{S}{0mm}{\vskip-0.5truein\centerline{\hskip-0.0truein\vbox{\psfig{figure=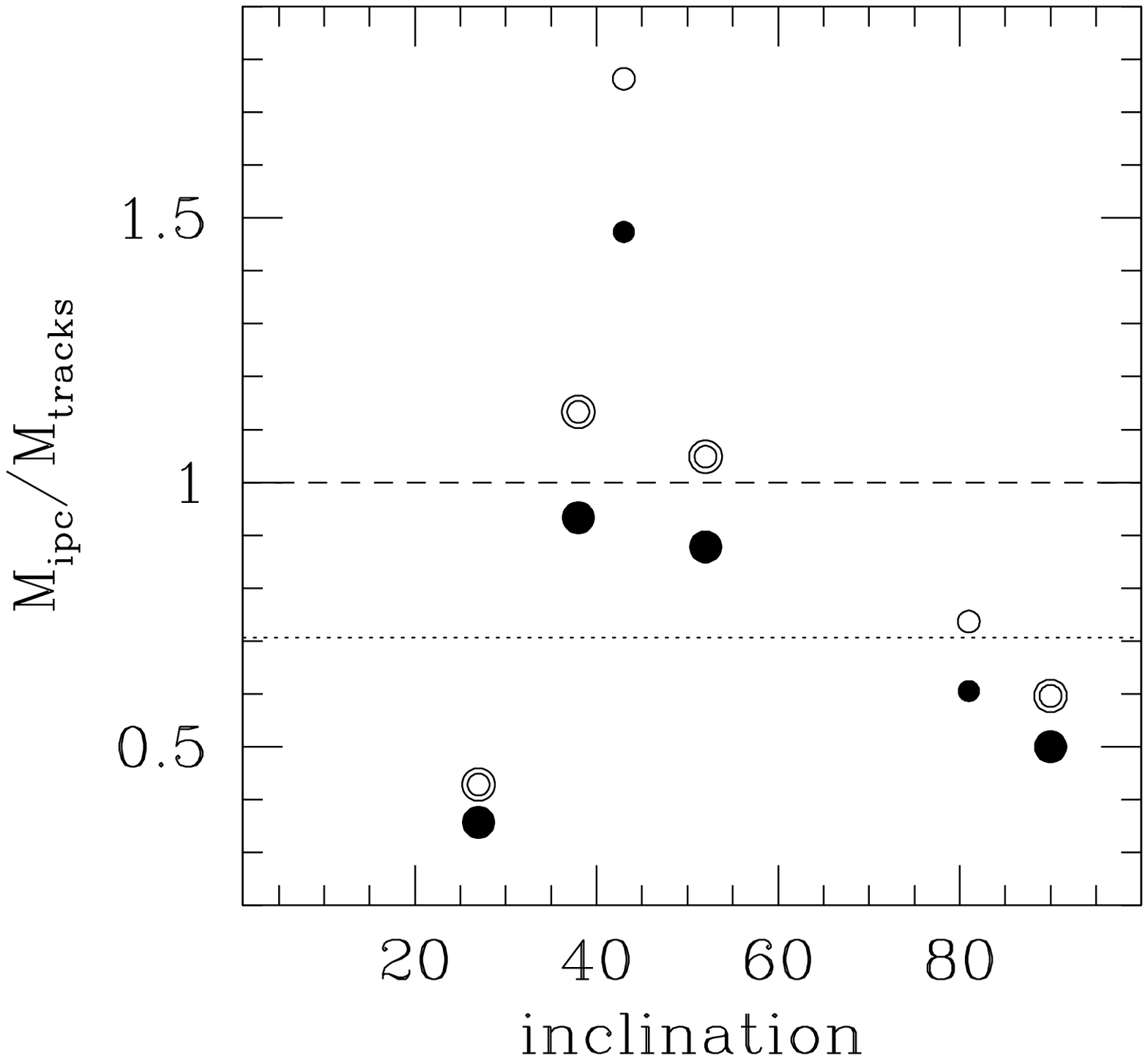,width=3.5truein,height=3.5truein,rwidth=3.25truein,rheight=3.25truein}}} 
\break\noindent
{\bf Figure 2.} The ratio of the dynamical mass estimate, \mipc, to
the mass derived from the PMS tracks, \mtra, is plotted against the
star's inclination to the line of sight (pole-on corresponds to 0
degrees).  The dynamical mass, \mipc, is calculated from the lower
characteristic velocity representing the deepest absorption and the
assumption that matter falls in from infinity (filled circles) or from five
stellar radii (open circles). The smaller symbols indicate an
uncertainty in the inclination of less than 10 degrees while the
larger symbols represent an uncertainty in the inclination of less
than 5 degrees. The highest point corresponds to DK Tau (a $2.5''$
binary). GI Tau has not been plotted because of the large difference
in radii between Hartigan \etal~(1995) and Gullbring \etal~(1998),
which leads to a large uncertainty in the inclination.}}

\tx The dynamical mass estimate of equation~(2) can also
tell us something about the geometry of the magnetospheric accretion
streams. The mass estimate uses the observed infall velocity along our
line of sight and so can constrain the orientation of the stream. Once
reliable masses are available from the PMS tracks we can combine the
mass estimate from the IPC velocity with the knowledge of the star's
inclination to our line of sight and constrain where on the star the
accretion stream impacts.  Figure~2 plots the ratio of the dynamical
mass estimate, \mipc\ (using the lower velocity corresponding to the
deepest part of the absorption profile), to the mass derived from the
PMS tracks, \mtra, against the inclination angle of the star for those
stars with both measured rotation periods and $v{\rm sin} i$
determinations (from Bouvier \etal~1995,).  The star GI Tau is
excluded from the plot because of the uncertainty in its radius
(between Hartigan \etal~[1995] and Gullbring \etal~[1998]) and hence
the inclination. The inclination angles are also listed in Table~1. GM
Aur is assigned an inclination of 90$\deg$ because the apparent ${\rm
sin} i > 1$.  From Figure~2, we see that stars oriented pole-on
($0\deg$ inclination) and equator-on ($90\deg$ inclination) have
dynamical mass estimates significantly lower than those derived from
the PMS tracks. In contrast, stars with moderate inclination angles
($30{\rm\ to\ }60\deg$) have dynamical mass estimates approximately
equal to or greater than those derived from PMS tracks.

Although the sample size is small, Figure~2 hints that the accretion
streams may impact preferentially at moderate latitudes.  This 
argues against scenarios of magnetospheric accretion where the
accretion streams are assumed to impact on, or near, the poles.  In
contrast, Muzerolle \etal~(1998) adopt a dipolar geometry where the
accretion stream impacts at moderate latitudes but that implies
small magnetospheric radii such that $\rmg \approx 2 \rs$. Kenyon et al.~1994
derive a similar model based on extensive multi-color observations of
the rotating ``hot spot'' observed on DR Tau.  A pure aligned dipolar
field implies a relation between the inclination at which the
accretion stream impacts ($\phi$, measured from the pole) and the
ratio of $\rmg/\rs$ given by (Hartmann \etal~1994)
$${\rmg\over\rs} = {\rm sin}^{-2} \phi.\eqno(4)$$ As the observed \vipc\
for the accretion stream should have a maximum where the star's
inclination $i$ is equal to this angle $\phi$, then we would expect
the inclination angles of 40 to 60 $\deg$ (as seen in Figure~2) to
correspond to disc truncation radii of $\rmg = 1.33 {\rm \ to\ } 2.4
\rs$, much smaller than the 5 \rst\ used here. These smaller values
would in turn imply significantly larger values of \mipc.  This
may imply that accretion geometry is complex, and not a pure, aligned
dipole.  Using the method described here, with a large enough sample,
it should be possible to constrain the geometry of the accretion
region.

\section{Discussion and Conclusions} 

\tx Determining accurate masses 
for large ensembles of pre-main-sequence (PMS) objects is 
presently impossible owing to our inability to adequately constrain the 
PMS evolutionary tracks. This limitation is a major obstacle
in determining the low-mass end ($0.05-0.5 \solm$)
of the IMF in regions of recent star formation (Hillenbrand~1997). 
Here, we have shown that lower limits on the stellar masses 
can be determined by measuring infall velocities on to these objects. 

In summary, the 
presence of red-shifted absorption profiles in the emission line
spectrum of PMS stars indicates near radial accretion.  In the
magnetospheric accretion model, the stellar magnetic field disrupts
the circumstellar disc at several stellar radii from where the matter
falls freely in along the field lines on to the stellar surface.  The
velocity of the infalling matter (from the IPC profile) measures the
potential energy at the stellar surface which, when combined with a
determination of the stellar radius, yields an estimate of the
stellar mass.  

For a collection of 13 stars (Hartigan \etal~1995), 11 have lower
limits which are comparable to or greater than the track masses and thus
can act as significant constraints on the stellar mass.  Using the
deepest part of the absorption profile as the impact velocity, and
assuming the disc is truncated at five stellar radii ($\rmg \approx 5
\rs$), 5 of the 13 stars have $\mip > \mtr$.  If the disc is truncated
at infinity, three of these five stars still have $\mip > \mtr$.  If
we model the impact velocity as the maximum velocity seen in
absorption, then 10 of the 13 stars have lower limits $\mip \simgreat
\mtr$, 8 of which have $\mip > \mtr$ if the disc is truncated at
infinity instead of at $5 \rs$.  Truncation radii smaller
than $5 \rs$ result in larger dynamical mass limits such
that if $\rmg \approx 2 \rs$, none of the lower limits would be
compatible with the PMS tracks.  From this we can conclude that i) the
method described here is a useful method for estimating PMS masses and
thus constraining the PMS evolutionary tracks, and ii) that there are
potentially significant problems in the DM94 tracks.

In addition, this 
method for setting lower limits on the stellar mass can also help
us constrain the geometry of the magnetospheric accretion.  Using
information on the star's inclination to our line of sight, we find 
that stars seen at moderate inclination appear to suffer less from 
projection effects. This implies that the accretion stream probably 
impacts on the star at moderate latitudes. 

Although this technique appears promising, 
additional work is required before quantitative tests of the available 
PMS tracks can be undertaken.  Specifically, we require; 
i) more extensive collections of IPC profiles, ii) improved estimates 
of stellar radii based on careful photometric and spectroscopic 
observations and iii) additional modelling efforts to determine 
the appropriate velocity to adopt from the IPC line profile, 
in order for this method to become a powerful 
tool in the constraint of the masses of PMS stars.  With these data, 
we can then begin to examine the PMS evolutionary tracks and study the geometry of the accretion streams.

\section*{Acknowledgments} 

\tx We thank Cathie Clarke, Matthew Bate,
Lee Hartmann, Jerome Bouvier and the referee for their comments. IAB
and CAT gratefully acknowledge support from PPARC advanced fellowships
KWS also acknowledges financial support from PPARC.  Support for MRM
was provided by NASA through Hubble Fellowship Grant \# HF1098.01
awarded by the Space Telescope Science Institute.  DFMF acknowledges
financial support from the ``Subprograma Ci\^encia e Tecnologia do
2$^o$ Quadro Comunit\'ario de Apoio''.

\section*{References} 

\bibitem Appenzeller I., Reitermann A., Stahl O., 1988, PASP, 100, 815

\bibitem Armitage P., Clarke C. J., 1996, MNRAS,  280, 458

\bibitem Armitage P., Clarke C. J., Tout C. A., 1998, MNRAS, submitted 

\bibitem Bouvier J., Covino E., Kovo O., Martin E. L., Matthews J. M., 
	Terranegra L., Beck S. C., 1995, A\&A, 299, 89

\bibitem Cameron A. C., Campbell C. G., 1993, A\&A, 274, 309

\bibitem Casey B. W., Mathieu R. D., Vaz L. P., Andersen J., Suntzeff N. B., 1998, AJ, in press

\bibitem D'Antona F., Mazzitelli I., 1994 ApJS, 90, 467 (DM94)

\bibitem Edwards S., 1997, IAUS, 182, 433

\bibitem Edwards S., Hartigan P., Ghandour L., Andrulis C., 1994, AJ, 108, 1056

\bibitem Edwards S., Strom S., Hartigan P., Strom K., Hillenbrand L., 
	Herbst W., Attridge J., Merill K., Probst R., Gatley I., 1993, AJ
	106, 372

\bibitem Folha D F. M., Emerson J. P, Calvet N, 1997, in poster proc., 
	IAU Symp. No. 182, eds. F. Malbet and A. Castets, p.272

\bibitem Folha D. F. M., Emerson J. P., 1998, in preparation

\bibitem Ghez A. M.,  Weinberger A. J., Neugebauer G., Matthews K., 
	McCarthy D. W., 1995,  AJ, 110, 753

\bibitem Ghez A. M., White R. J., Simon M., 1997, ApJ, 490, 353

\bibitem Gullbring E., Hartmann L., Briceno C., Calvet N., 1998, ApJ, 492, 323

\bibitem Hartigan P., Edwards S., Ghandour L., 1995, ApJ, 452, 736

\bibitem Hartmann L., Cassen P. Kenyon S., 1997, ApJ, 475, 770

\bibitem Hartmann L., Hewett R., Calvet N., 1994, ApJ, 42, 669

\bibitem Hillenbrand L. A., 1997, AJ, 113, 1733

\bibitem Kenyon S. J., Hartmann L., 1995, ApJS, 101, 117

\bibitem Kenyon S. J., Yi I., Hartmann L., 1996, ApJ, 462, 439

\bibitem Koerner, D., Sargent, A., and Beckwith, S.V.W., 1993, Icarus, 106, 2

\bibitem Konigl A., 1991, ApJL, 370, L39 

\bibitem Kroupa P., Tout C. A., and Gilmore, G., 1993, MNRAS, 262, 545

\bibitem Lee, C.W., 1992, Ph.D. Thesis, U. Wisconsin 

\bibitem Lin D. N., Pringle J. E., 1990, ApJ, 358, 515

\bibitem Luhmann K. L., Rieke G. H., 1998, preprint

\bibitem Mathieu R. D., 1994, ARA\&A, 32, 465

\bibitem Meyer M. R., Calvet N., Hillenbrand L. A., 1997, AJ, 114, 288

\bibitem Muzerolle J., Calvet N., Hartmann L., 1998, ApJ, 492, 743

\bibitem Nakajima T., Oppenheimer B. R., Kulkarni S. R., Golimowski D. A., 
	Matthews K., Durrance S. T., 1995, Nature, 378, 463

\bibitem Popper D. M., 1980, ARA\&A, 18, 115

\bibitem Pringle J. E., 1981, ARA\&A, 19, 137

\bibitem Rebolo R., Zapatero-Osorio M. R., Martin E. L., 1995, Nature, 377, 129

\bibitem Seis L., Forestini M., Bertout C., 1997, A\&A, 326, 101

\bibitem Simon M., Prato L., 1995, ApJ, 450, 824

\bibitem Smith K. W., Bonnell I. A., Lewis G. F., Bunclark P. S., 1997, 
	MNRAS, 289, 151

\bibitem Swenson F. J., Faulkner J., Rogers F. J., Iglesias C. A., 1994, 
	ApJ, 425, 286

\bibitem Walker M. F., 1978, ApJ, 224, 546

\vfill\eject\end

%% file: mnd.tex

\catcode `\@=11 

\def\@version{1.3}
\def\@verdate{28.11.1992}


%
%
%
%
%
%

\font\fiverm=cmr5
\font\fivei=cmmi5	\skewchar\fivei='177
\font\fivesy=cmsy5	\skewchar\fivesy='60
\font\fivebf=cmbx5

\font\sevenrm=cmr7
\font\seveni=cmmi7	\skewchar\seveni='177
\font\sevensy=cmsy7	\skewchar\sevensy='60
\font\sevenbf=cmbx7

\font\eightrm=cmr8
\font\eightbf=cmbx8
\font\eightit=cmti8
\font\eighti=cmmi8			\skewchar\eighti='177
\font\eightmib=cmmib10 at 8pt	\skewchar\eightmib='177
\font\eightsy=cmsy8			\skewchar\eightsy='60
\font\eightsyb=cmbsy10 at 8pt	\skewchar\eightsyb='60
\font\eightsl=cmsl8
\font\eighttt=cmtt8			\hyphenchar\eighttt=-1
\font\eightcsc=cmcsc10 at 8pt
\font\eightsf=cmss8

\font\ninerm=cmr9
\font\ninebf=cmbx9
\font\nineit=cmti9
\font\ninei=cmmi9			\skewchar\ninei='177
\font\ninemib=cmmib10 at 9pt	\skewchar\ninemib='177
\font\ninesy=cmsy9			\skewchar\ninesy='60
\font\ninesyb=cmbsy10 at 9pt	\skewchar\ninesyb='60
\font\ninesl=cmsl9
\font\ninett=cmtt9			\hyphenchar\ninett=-1
\font\ninecsc=cmcsc10 at 9pt
\font\ninesf=cmss9

\font\tenrm=cmr10
\font\tenbf=cmbx10
\font\tenit=cmti10
\font\teni=cmmi10		\skewchar\teni='177
\font\tenmib=cmmib10	\skewchar\tenmib='177
\font\tensy=cmsy10		\skewchar\tensy='60
\font\tensyb=cmbsy10	\skewchar\tensyb='60
\font\tenex=cmex10
\font\tensl=cmsl10
\font\tentt=cmtt10		\hyphenchar\tentt=-1
\font\tencsc=cmcsc10
\font\tensf=cmss10

\font\elevenrm=cmr10 scaled \magstephalf
\font\elevenbf=cmbx10 scaled \magstephalf
\font\elevenit=cmti10 scaled \magstephalf
\font\eleveni=cmmi10 scaled \magstephalf	\skewchar\eleveni='177
\font\elevenmib=cmmib10 scaled \magstephalf	\skewchar\elevenmib='177
\font\elevensy=cmsy10 scaled \magstephalf	\skewchar\elevensy='60
\font\elevensyb=cmbsy10 scaled \magstephalf	\skewchar\elevensyb='60
\font\elevensl=cmsl10 scaled \magstephalf
\font\eleventt=cmtt10 scaled \magstephalf	\hyphenchar\eleventt=-1
\font\elevencsc=cmcsc10 scaled \magstephalf
\font\elevensf=cmss10 scaled \magstephalf

\font\fourteenrm=cmr10 scaled \magstep2
\font\fourteenbf=cmbx10 scaled \magstep2
\font\fourteenit=cmti10 scaled \magstep2
\font\fourteeni=cmmi10 scaled \magstep2		\skewchar\fourteeni='177
\font\fourteenmib=cmmib10 scaled \magstep2	\skewchar\fourteenmib='177
\font\fourteensy=cmsy10 scaled \magstep2	\skewchar\fourteensy='60
\font\fourteensyb=cmbsy10 scaled \magstep2	\skewchar\fourteensyb='60
\font\fourteensl=cmsl10 scaled \magstep2
\font\fourteentt=cmtt10 scaled \magstep2	\hyphenchar\fourteentt=-1
\font\fourteencsc=cmcsc10 scaled \magstep2
\font\fourteensf=cmss10 scaled \magstep2

\font\seventeenrm=cmr10 scaled \magstep3
\font\seventeenbf=cmbx10 scaled \magstep3
\font\seventeenit=cmti10 scaled \magstep3
\font\seventeeni=cmmi10 scaled \magstep3	\skewchar\seventeeni='177
\font\seventeenmib=cmmib10 scaled \magstep3	\skewchar\seventeenmib='177
\font\seventeensy=cmsy10 scaled \magstep3	\skewchar\seventeensy='60
\font\seventeensyb=cmbsy10 scaled \magstep3	\skewchar\seventeensyb='60
\font\seventeensl=cmsl10 scaled \magstep3
\font\seventeentt=cmtt10 scaled \magstep3	\hyphenchar\seventeentt=-1
\font\seventeencsc=cmcsc10 scaled \magstep3
\font\seventeensf=cmss10 scaled \magstep3

\def\@typeface{Computer Modern} 

\def\hexnumber@#1{\ifnum#1<10 \number#1\else
 \ifnum#1=10 A\else\ifnum#1=11 B\else\ifnum#1=12 C\else
 \ifnum#1=13 D\else\ifnum#1=14 E\else\ifnum#1=15 F\fi\fi\fi\fi\fi\fi\fi}

\def\mib{\hexnumber@\mibfam}
\def\syb{\hexnumber@\sybfam}

\def\makestrut{%
  \setbox\strutbox=\hbox{%
    \vrule height.7\baselineskip depth.3\baselineskip width 0pt}%
}

\def\bls#1{%
  \normalbaselineskip=#1%
  \normalbaselines%
  \makestrut%
}

%

\newfam\mibfam 
\newfam\sybfam 
\newfam\scfam  
\newfam\sffam  

\def\em{\ifdim\fontdimen1\font>0 \rm\else\it\fi}

\textfont3=\tenex
\scriptfont3=\tenex
\scriptscriptfont3=\tenex

\def\eightpoint{
  \def\rm{\fam0\eightrm}%
  \textfont0=\eightrm \scriptfont0=\sevenrm \scriptscriptfont0=\fiverm%
  \textfont1=\eighti  \scriptfont1=\seveni  \scriptscriptfont1=\fivei%
  \textfont2=\eightsy \scriptfont2=\sevensy \scriptscriptfont2=\fivesy%
  \textfont\itfam=\eightit\def\it{\fam\itfam\eightit}%
  \textfont\bffam=\eightbf%
    \scriptfont\bffam=\sevenbf%
      \scriptscriptfont\bffam=\fivebf%
  \def\bf{\fam\bffam\eightbf}%
  \textfont\slfam=\eightsl\def\sl{\fam\slfam\eightsl}%
  \textfont\ttfam=\eighttt\def\tt{\fam\ttfam\eighttt}%
  \textfont\scfam=\eightcsc\def\sc{\fam\scfam\eightcsc}%
  \textfont\sffam=\eightsf\def\sf{\fam\sffam\eightsf}%
  \textfont\mibfam=\eightmib%
  \textfont\sybfam=\eightsyb%
  \bls{10pt}%
}

\def\ninepoint{
  \def\rm{\fam0\ninerm}%
  \textfont0=\ninerm \scriptfont0=\sevenrm \scriptscriptfont0=\fiverm%
  \textfont1=\ninei  \scriptfont1=\seveni  \scriptscriptfont1=\fivei%
  \textfont2=\ninesy \scriptfont2=\sevensy \scriptscriptfont2=\fivesy%
  \textfont\itfam=\nineit\def\it{\fam\itfam\nineit}%
  \textfont\bffam=\ninebf%
    \scriptfont\bffam=\sevenbf%
      \scriptscriptfont\bffam=\fivebf%
  \def\bf{\fam\bffam\ninebf}%
  \textfont\slfam=\ninesl\def\sl{\fam\slfam\ninesl}%
  \textfont\ttfam=\ninett\def\tt{\fam\ttfam\ninett}%
  \textfont\scfam=\ninecsc\def\sc{\fam\scfam\ninecsc}%
  \textfont\sffam=\ninesf\def\sf{\fam\sffam\ninesf}%
  \textfont\mibfam=\ninemib%
  \textfont\sybfam=\ninesyb%
  \bls{12pt}%
}

\def\tenpoint{
  \def\rm{\fam0\tenrm}%
  \textfont0=\tenrm \scriptfont0=\sevenrm \scriptscriptfont0=\fiverm%
  \textfont1=\teni  \scriptfont1=\seveni  \scriptscriptfont1=\fivei%
  \textfont2=\tensy \scriptfont2=\sevensy \scriptscriptfont2=\fivesy%
  \textfont\itfam=\tenit\def\it{\fam\itfam\tenit}%
  \textfont\bffam=\tenbf%
    \scriptfont\bffam=\sevenbf%
      \scriptscriptfont\bffam=\fivebf%
  \def\bf{\fam\bffam\tenbf}%
  \textfont\slfam=\tensl\def\sl{\fam\slfam\tensl}%
  \textfont\ttfam=\tentt\def\tt{\fam\ttfam\tentt}%
  \textfont\scfam=\tencsc\def\sc{\fam\scfam\tencsc}%
  \textfont\sffam=\tensf\def\sf{\fam\sffam\tensf}%
  \textfont\mibfam=\tenmib%
  \textfont\sybfam=\tensyb%
  \bls{12pt}%
}

\def\elevenpoint{
  \def\rm{\fam0\elevenrm}%
  \textfont0=\elevenrm \scriptfont0=\eightrm \scriptscriptfont0=\fiverm%
  \textfont1=\eleveni  \scriptfont1=\eighti  \scriptscriptfont1=\fivei%
  \textfont2=\elevensy \scriptfont2=\eightsy \scriptscriptfont2=\fivesy%
  \textfont\itfam=\elevenit\def\it{\fam\itfam\elevenit}%
  \textfont\bffam=\elevenbf%
    \scriptfont\bffam=\eightbf%
      \scriptscriptfont\bffam=\fivebf%
  \def\bf{\fam\bffam\elevenbf}%
  \textfont\slfam=\elevensl\def\sl{\fam\slfam\elevensl}%
  \textfont\ttfam=\eleventt\def\tt{\fam\ttfam\eleventt}%
  \textfont\scfam=\elevencsc\def\sc{\fam\scfam\elevencsc}%
  \textfont\sffam=\elevensf\def\sf{\fam\sffam\elevensf}%
  \textfont\mibfam=\elevenmib%
  \textfont\sybfam=\elevensyb%
  \bls{13pt}%
}

\def\fourteenpoint{
  \def\rm{\fam0\fourteenrm}%
  \textfont0\fourteenrm  \scriptfont0\tenrm  \scriptscriptfont0\sevenrm%
  \textfont1\fourteeni   \scriptfont1\teni   \scriptscriptfont1\seveni%
  \textfont2\fourteensy  \scriptfont2\tensy  \scriptscriptfont2\sevensy%
  \textfont\itfam=\fourteenit\def\it{\fam\itfam\fourteenit}%
  \textfont\bffam=\fourteenbf%
    \scriptfont\bffam=\tenbf%
      \scriptscriptfont\bffam=\sevenbf%
  \def\bf{\fam\bffam\fourteenbf}%
  \textfont\slfam=\fourteensl\def\sl{\fam\slfam\fourteensl}%
  \textfont\ttfam=\fourteentt\def\tt{\fam\ttfam\fourteentt}%
  \textfont\scfam=\fourteencsc\def\sc{\fam\scfam\fourteencsc}%
  \textfont\sffam=\fourteensf\def\sf{\fam\sffam\fourteensf}%
  \textfont\mibfam=\fourteenmib%
  \textfont\sybfam=\fourteensyb%
  \bls{17pt}%
}

\def\seventeenpoint{
  \def\rm{\fam0\seventeenrm}%
  \textfont0\seventeenrm  \scriptfont0\elevenrm  \scriptscriptfont0\ninerm%
  \textfont1\seventeeni   \scriptfont1\eleveni   \scriptscriptfont1\ninei%
  \textfont2\seventeensy  \scriptfont2\elevensy  \scriptscriptfont2\ninesy%
  \textfont\itfam=\seventeenit\def\it{\fam\itfam\seventeenit}%
  \textfont\bffam=\seventeenbf%
    \scriptfont\bffam=\elevenbf%
      \scriptscriptfont\bffam=\ninebf%
  \def\bf{\fam\bffam\seventeenbf}%
  \textfont\slfam=\seventeensl\def\sl{\fam\slfam\seventeensl}%
  \textfont\ttfam=\seventeentt\def\tt{\fam\ttfam\seventeentt}%
  \textfont\scfam=\seventeencsc\def\sc{\fam\scfam\seventeencsc}%
  \textfont\sffam=\seventeensf\def\sf{\fam\sffam\seventeensf}%
  \textfont\mibfam=\seventeenmib%
  \textfont\sybfam=\seventeensyb%
  \bls{20pt}%
}

\lineskip=1pt      \normallineskip=\lineskip
\lineskiplimit=0pt \normallineskiplimit=\lineskiplimit




\def\Nulle{0}  
\def\Aue{1}    
\def\Afe{2}    
\def\Ace{3}    
\def\Sue{4}    
\def\Hae{5}    
\def\Hbe{6}    
\def\Hce{7}    
\def\Hde{8}    
\def\Kwe{9}    
\def\Txe{10}   
\def\Lie{11}   
\def\Bbe{12}   


\newdimen\DimenA
\newbox\BoxA

\newcount\LastMac \LastMac=\Nulle
\newcount\HeaderNumber \HeaderNumber=0
\newcount\DefaultHeader \DefaultHeader=\HeaderNumber
\newskip\Indent

\newskip\half      \half=5.5pt plus 1.5pt minus 2.25pt
\newskip\one       \one=11pt plus 3pt minus 5.5pt
\newskip\onehalf   \onehalf=16.5pt plus 5.5pt minus 8.25pt
\newskip\two       \two=22pt plus 5.5pt minus 11pt

\def\Half{\vskip-\lastskip\vskip\half}
\def\One{\vskip-\lastskip\vskip\one}
\def\OneHalf{\vskip-\lastskip\vskip\onehalf}
\def\Two{\vskip-\lastskip\vskip\two}


\def\rTenPT{10pt plus \Feathering}

\def\TenPT{10pt plus \Feathering} 
\def\ElevenPT{11pt plus \Feathering}

\def\Raggedright{
 \rightskip=0pt plus \hsize
}

\def\Fullout{
\rightskip=0pt
}

\def\Hang#1#2{
 \hangindent=#1
 \hangafter=#2
}

\def\EveryMac{
 \Fullout
 \everypar{}
}



\def\title#1{
 \EveryMac
 \LastMac=\Nulle
 \global\HeaderNumber=0
 \global\DefaultHeader=1
 \vbox to 1pc{\vss}
 \seventeenpoint
 \Raggedright
 \noindent \bf #1
}

\def\author#1{
 \EveryMac
 \ifnum\LastMac=\Afe \OneHalf
  \else \Two
 \fi
 \LastMac=\Aue
 \fourteenpoint
 \Raggedright
 \noindent \rm #1\par
 \vskip 3pt\relax
}

\def\affiliation#1{
 \EveryMac
 \LastMac=\Afe
 \eightpoint\bls{\TenPT}
 \Raggedright
 \noindent \it #1\par
}

\def\acceptedline#1{
 \EveryMac
 \Two
 \LastMac=\Ace
 \eightpoint\bls{\TenPT}
 \Raggedright
 \noindent \rm #1
}

\def\abstract{%
 \EveryMac
 \Two
 \LastMac=\Sue
 \everypar{\Hang{11pc}{0}}
 \noindent\ninebf ABSTRACT\par
 \tenpoint\bls{\ElevenPT}
 \Fullout
 \noindent\rm
}

\def\keywords{
 \EveryMac
 \Half
 \LastMac=\Kwe
 \everypar{\Hang{11pc}{0}}
 \tenpoint\bls{\ElevenPT}
 \Fullout
 \noindent\hbox{\bf Key words:\ }
 \rm
}


\def\maketitle{%
  \Two%
  \EndOpening%
  \MakePage%
}


\def\pageoffset#1#2{\hoffset=#1\relax\voffset=#2\relax}


\def\Autonumber{
 \global\AutoNumbertrue  
}

\newif\ifAutoNumber \AutoNumberfalse
\newcount\Sec        
\newcount\SecSec
\newcount\SecSecSec

\Sec=0

\def\:{\let\@sptoken= } \:  
\def\:{\@xifnch} \expandafter\def\: {\futurelet\@tempc\@ifnch}

\def\@ifnextchar#1#2#3{%
  \let\@tempMACe #1%
  \def\@tempMACa{#2}%
  \def\@tempMACb{#3}%
  \futurelet \@tempMACc\@ifnch%
}

\def\@ifnch{%
\ifx \@tempMACc \@sptoken%
  \let\@tempMACd\@xifnch%
\else%
  \ifx \@tempMACc \@tempMACe%
    \let\@tempMACd\@tempMACa%
  \else%
    \let\@tempMACd\@tempMACb%
  \fi%
\fi%
\@tempMACd%
}

\def\@ifstar#1#2{\@ifnextchar *{\def\@tempMACa*{#1}\@tempMACa}{#2}}

\def\section{\@ifstar{\@ssection}{\@section}}

\def\@section#1{
 \EveryMac
 \Two
 \LastMac=\Hae
 \ninepoint\bls{\ElevenPT}
 \bf
 \Raggedright
 \ifAutoNumber
  \advance\Sec by 1
  \noindent\number\Sec\hskip 1pc \uppercase{#1}
  \SecSec=0
 \else
  \noindent \uppercase{#1}
 \fi
 \nobreak
}

\def\@ssection#1{
 \EveryMac
 \ifnum\LastMac=\Hae \Half
  \else \OneHalf
 \fi
 \LastMac=\Hae
 \tenpoint\bls{\ElevenPT}
 \bf
 \Raggedright
 \noindent\uppercase{#1}
}

\def\subsection#1{
 \EveryMac
 \ifnum\LastMac=\Hae \Half
  \else \OneHalf
 \fi
 \LastMac=\Hbe
 \tenpoint\bls{\ElevenPT}
 \bf
 \Raggedright
 \ifAutoNumber
  \advance\SecSec by 1
  \noindent\number\Sec.\number\SecSec
  \hskip 1pc #1
  \SecSecSec=0
 \else
  \noindent #1
 \fi
 \nobreak
}

\def\subsubsection#1{
 \EveryMac
 \ifnum\LastMac=\Hbe \Half
  \else \OneHalf
 \fi
 \LastMac=\Hce
 \ninepoint\bls{\ElevenPT}
 \it
 \Raggedright
 \ifAutoNumber
  \advance\SecSecSec by 1
  \noindent\number\Sec.\number\SecSec.\number\SecSecSec
  \hskip 1pc #1
 \else
  \noindent #1
 \fi
 \nobreak
}

\def\paragraph#1{
 \EveryMac
 \One
 \LastMac=\Hde
 \ninepoint\bls{\ElevenPT}
 \noindent \it #1
 \rm
}


\def\tx{
 \EveryMac
 \ifnum\LastMac=\Lie \Half\fi
 \ifnum\LastMac=\Hae \nobreak\Half\fi
 \ifnum\LastMac=\Hbe \nobreak\Half\fi
 \ifnum\LastMac=\Hce \nobreak\Half\fi
 \ifnum\LastMac=\Lie \else \noindent\fi
 \LastMac=\Txe
 \ninepoint\bls{\ElevenPT}
 \rm
}


\def\item{
 \par
 \EveryMac
 \ifnum\LastMac=\Lie
  \else \Half
 \fi
 \LastMac=\Lie
 \ninepoint\bls{\ElevenPT}
 \rm
}


\def\bibitem{
 \par
 \EveryMac
 \ifnum\LastMac=\Bbe
  \else \Half
 \fi
 \LastMac=\Bbe
 \Hang{1.5em}{1}
 \eightpoint\bls{\TenPT}
 \Raggedright
 \noindent \rm
}


\newtoks\CatchLine

\def\@journal{Mon.\ Not.\ R.\ Astron.\ Soc.\ }  
\def\@pubyear{1993}        
\def\@pagerange{000--000}  
\def\@volume{000}          
\def\@microfiche{}         %

\def\pubyear#1{\gdef\@pubyear{#1}\@makecatchline}
\def\pagerange#1{\gdef\@pagerange{#1}\@makecatchline}
\def\volume#1{\gdef\@volume{#1}\@makecatchline}
\def\microfiche#1{\gdef\@microfiche{and Microfiche\ #1}\@makecatchline}

\def\@makecatchline{%
  \global\CatchLine{%
    {\rm \@journal {\bf \@volume},\ \@pagerange\ (\@pubyear)\ \@microfiche}}%
}

\@makecatchline 

\newtoks\LeftHeader
\def\shortauthor#1{
 \global\LeftHeader{#1}
}

\newtoks\RightHeader
\def\shorttitle#1{
 \global\RightHeader{#1}
}

\def\PageHead{
 \EveryMac
 \ifnum\HeaderNumber=1 \Pagehead
  \else \Catchline
 \fi
}

\def\Catchline{%
 \vbox to 0pt{\vskip-22.5pt
  \hbox to \PageWidth{\vbox to8.5pt{}\noindent
  \eightpoint\the\CatchLine\hfill}\vss}
 \nointerlineskip
}

\def\Pagehead{%
 \ifodd\pageno
   \vbox to 0pt{\vskip-22.5pt
   \hbox to \PageWidth{\vbox to8.5pt{}\elevenpoint\it\noindent
    \hfill\the\RightHeader\hskip1.5em\rm\folio}\vss}
 \else
   \vbox to 0pt{\vskip-22.5pt
   \hbox to \PageWidth{\vbox to8.5pt{}\elevenpoint\rm\noindent
   \folio\hskip1.5em\it\the\LeftHeader\hfill}\vss}
 \fi
 \nointerlineskip
}

\def\PageFoot{} 

\def\authorcomment#1{%
  \gdef\PageFoot{%
    \nointerlineskip%
    \vbox to 22pt{\vfil%
      \hbox to \PageWidth{\elevenpoint\rm\noindent \hfil #1 \hfil}}%
  }%
}

\everydisplay{\displaysetup}

\newif\ifeqno
\newif\ifleqno

\def\displaysetup#1$${%
 \displaytest#1\eqno\eqno\displaytest
}

\def\displaytest#1\eqno#2\eqno#3\displaytest{%
 \if!#3!\ldisplaytest#1\leqno\leqno\ldisplaytest
 \else\eqnotrue\leqnofalse\def\eqn{#2}\def\eq{#1}\fi
 \generaldisplay$$}

\def\ldisplaytest#1\leqno#2\leqno#3\ldisplaytest{%
 \def\eq{#1}%
 \if!#3!\eqnofalse\else\eqnotrue\leqnotrue
  \def\eqn{#2}\fi}

\def\generaldisplay{%
\ifeqno \ifleqno 
   \hbox to \hsize{\noindent
     $\displaystyle\eq$\hfil$\displaystyle\eqn$}
  \else
    \hbox to \hsize{\noindent
     $\displaystyle\eq$\hfil$\displaystyle\eqn$}
  \fi
 \else
 \hbox to \hsize{\vbox{\noindent
  $\displaystyle\eq$\hfil}}
 \fi
}

\def\@notice{%
  \par\Two%
  \bls{12pt}%
  \noindent\tenrm This paper has been produced using the Blackwell
                  Scientific Publications \TeX\ macros.%
}

\outer\def\bye{\@notice\par\vfill\supereject\end}

\everyjob{%
  \Warn{Monthly notices of the RAS journal style (\@typeface)\space
        v\@version,\space \@verdate.}\Warn{}%
}




\newif\if@debug \@debugfalse  

\def\Print#1{\if@debug\immediate\write16{#1}\else \fi}
\def\Warn#1{\immediate\write16{#1}}
\def\wlog#1{}

\newcount\Iteration 

\newif\ifFigureBoxes  
\FigureBoxestrue

\def\Single{0} \def\Double{1}                 
\def\Figure{0} \def\Table{1}                  

\def\InStack{0}  
\def\InZoneA{1}
\def\InZoneB{2}
\def\InZoneC{3}

\newcount\TEMPCOUNT 
\newdimen\TEMPDIMEN 
\newbox\TEMPBOX     
\newbox\VOIDBOX     

\newcount\LengthOfStack 
\newcount\MaxItems      
\newcount\StackPointer
\newcount\Point         
\newcount\NextFigure    
\newcount\NextTable     
\newcount\NextItem      

\newcount\StatusStack   
\newcount\NumStack      
\newcount\TypeStack     
\newcount\SpanStack     
\newcount\BoxStack      

\newcount\ItemSTATUS    
\newcount\ItemNUMBER    
\newcount\ItemTYPE      
\newcount\ItemSPAN      
\newbox\ItemBOX         
\newdimen\ItemSIZE      

\newdimen\PageHeight    
\newdimen\TextLeading   
\newdimen\Feathering    
\newcount\LinesPerPage  
\newdimen\ColumnWidth   
\newdimen\ColumnGap     
\newdimen\PageWidth     
\newdimen\BodgeHeight   
\newcount\Leading       

\newdimen\ZoneBSize  
\newdimen\TextSize   
\newbox\ZoneABOX     
\newbox\ZoneBBOX     
\newbox\ZoneCBOX     

\newif\ifFirstSingleItem
\newif\ifFirstZoneA
\newif\ifMakePageInComplete
\newif\ifMoreFigures \MoreFiguresfalse 
\newif\ifMoreTables  \MoreTablesfalse  

\newif\ifFigInZoneB 
\newif\ifFigInZoneC 
\newif\ifTabInZoneB 
\newif\ifTabInZoneC

\newif\ifZoneAFullPage

\newbox\MidBOX    
\newbox\LeftBOX
\newbox\RightBOX
\newbox\PageBOX   

\newif\ifLeftCOL  
\LeftCOLtrue

\newdimen\ZoneBAdjust

\newcount\ItemFits
\def\Yes{1}
\def\No{2}




\MaxItems=15
\NextFigure=0        
\NextTable=1

\BodgeHeight=6pt
\TextLeading=11pt    
\Leading=11
\Feathering=0pt      
\LinesPerPage=61     
\topskip=\TextLeading
\ColumnWidth=20pc    
\ColumnGap=2pc       

\def\ItemSep{\vskip \TextLeading plus \TextLeading minus 4pt}

\FigureBoxesfalse 

\parskip=0pt
\parindent=18pt
\widowpenalty=0
\clubpenalty=10000
\tolerance=1500
\hbadness=1500
\abovedisplayskip=6pt plus 2pt minus 2pt
\belowdisplayskip=6pt plus 2pt minus 2pt
\abovedisplayshortskip=6pt plus 2pt minus 2pt
\belowdisplayshortskip=6pt plus 2pt minus 2pt

\PageHeight=\TextLeading 
\multiply\PageHeight by \LinesPerPage
\advance\PageHeight by \topskip

\PageWidth=2\ColumnWidth
\advance\PageWidth by \ColumnGap




\newcount\DUMMY \StatusStack=\allocationnumber
\newcount\DUMMY \newcount\DUMMY \newcount\DUMMY 
\newcount\DUMMY \newcount\DUMMY \newcount\DUMMY 
\newcount\DUMMY \newcount\DUMMY \newcount\DUMMY
\newcount\DUMMY \newcount\DUMMY \newcount\DUMMY 
\newcount\DUMMY \newcount\DUMMY \newcount\DUMMY

\newcount\DUMMY \NumStack=\allocationnumber
\newcount\DUMMY \newcount\DUMMY \newcount\DUMMY 
\newcount\DUMMY \newcount\DUMMY \newcount\DUMMY 
\newcount\DUMMY \newcount\DUMMY \newcount\DUMMY 
\newcount\DUMMY \newcount\DUMMY \newcount\DUMMY 
\newcount\DUMMY \newcount\DUMMY \newcount\DUMMY

\newcount\DUMMY \TypeStack=\allocationnumber
\newcount\DUMMY \newcount\DUMMY \newcount\DUMMY 
\newcount\DUMMY \newcount\DUMMY \newcount\DUMMY 
\newcount\DUMMY \newcount\DUMMY \newcount\DUMMY 
\newcount\DUMMY \newcount\DUMMY \newcount\DUMMY 
\newcount\DUMMY \newcount\DUMMY \newcount\DUMMY

\newcount\DUMMY \SpanStack=\allocationnumber
\newcount\DUMMY \newcount\DUMMY \newcount\DUMMY 
\newcount\DUMMY \newcount\DUMMY \newcount\DUMMY 
\newcount\DUMMY \newcount\DUMMY \newcount\DUMMY 
\newcount\DUMMY \newcount\DUMMY \newcount\DUMMY 
\newcount\DUMMY \newcount\DUMMY \newcount\DUMMY

\newbox\DUMMY   \BoxStack=\allocationnumber
\newbox\DUMMY   \newbox\DUMMY \newbox\DUMMY 
\newbox\DUMMY   \newbox\DUMMY \newbox\DUMMY 
\newbox\DUMMY   \newbox\DUMMY \newbox\DUMMY 
\newbox\DUMMY   \newbox\DUMMY \newbox\DUMMY 
\newbox\DUMMY   \newbox\DUMMY \newbox\DUMMY

\def\wlog{\immediate\write-1}


\def\GetItemAll#1{%
 \GetItemSTATUS{#1}
 \GetItemNUMBER{#1}
 \GetItemTYPE{#1}
 \GetItemSPAN{#1}
 \GetItemBOX{#1}
}

\def\GetItemSTATUS#1{%
 \Point=\StatusStack
 \advance\Point by #1
 \global\ItemSTATUS=\count\Point
}

\def\GetItemNUMBER#1{%
 \Point=\NumStack
 \advance\Point by #1
 \global\ItemNUMBER=\count\Point
}

\def\GetItemTYPE#1{%
 \Point=\TypeStack
 \advance\Point by #1
 \global\ItemTYPE=\count\Point
}

\def\GetItemSPAN#1{%
 \Point\SpanStack
 \advance\Point by #1
 \global\ItemSPAN=\count\Point
}

\def\GetItemBOX#1{%
 \Point=\BoxStack
 \advance\Point by #1
 \global\setbox\ItemBOX=\vbox{\copy\Point}
 \global\ItemSIZE=\ht\ItemBOX
 \global\advance\ItemSIZE by \dp\ItemBOX
 \TEMPCOUNT=\ItemSIZE
 \divide\TEMPCOUNT by \Leading
 \divide\TEMPCOUNT by 65536
 \advance\TEMPCOUNT by 1
 \ItemSIZE=\TEMPCOUNT pt
 \global\multiply\ItemSIZE by \Leading
}


\def\JoinStack{%
 \ifnum\LengthOfStack=\MaxItems 
  \Warn{WARNING: Stack is full...some items will be lost!}
 \else
  \Point=\StatusStack
  \advance\Point by \LengthOfStack
  \global\count\Point=\ItemSTATUS
  \Point=\NumStack
  \advance\Point by \LengthOfStack
  \global\count\Point=\ItemNUMBER
  \Point=\TypeStack
  \advance\Point by \LengthOfStack
  \global\count\Point=\ItemTYPE
  \Point\SpanStack
  \advance\Point by \LengthOfStack
  \global\count\Point=\ItemSPAN
  \Point=\BoxStack
  \advance\Point by \LengthOfStack
  \global\setbox\Point=\vbox{\copy\ItemBOX}
  \global\advance\LengthOfStack by 1
  \ifnum\ItemTYPE=\Figure 
   \global\MoreFigurestrue
  \else
   \global\MoreTablestrue
  \fi
 \fi
}


\def\LeaveStack#1{%
 {\Iteration=#1
 \loop
 \ifnum\Iteration<\LengthOfStack
  \advance\Iteration by 1
  \GetItemSTATUS{\Iteration}
   \advance\Point by -1
   \global\count\Point=\ItemSTATUS
  \GetItemNUMBER{\Iteration}
   \advance\Point by -1
   \global\count\Point=\ItemNUMBER
  \GetItemTYPE{\Iteration}
   \advance\Point by -1
   \global\count\Point=\ItemTYPE
  \GetItemSPAN{\Iteration}
   \advance\Point by -1
   \global\count\Point=\ItemSPAN
  \GetItemBOX{\Iteration}
   \advance\Point by -1
   \global\setbox\Point=\vbox{\copy\ItemBOX}
 \repeat}
 \global\advance\LengthOfStack by -1
}


\newif\ifStackNotClean

\def\CleanStack{%
 \StackNotCleantrue
 {\Iteration=0
  \loop
   \ifStackNotClean
    \GetItemSTATUS{\Iteration}
    \ifnum\ItemSTATUS=\InStack
     \advance\Iteration by 1
     \else
      \LeaveStack{\Iteration}
    \fi
   \ifnum\LengthOfStack<\Iteration
    \StackNotCleanfalse
   \fi
 \repeat}
}


\def\FindItem#1#2{%
 \global\StackPointer=-1 
 {\Iteration=0
  \loop
  \ifnum\Iteration<\LengthOfStack
   \GetItemSTATUS{\Iteration}
   \ifnum\ItemSTATUS=\InStack
    \GetItemTYPE{\Iteration}
    \ifnum\ItemTYPE=#1
     \GetItemNUMBER{\Iteration}
     \ifnum\ItemNUMBER=#2
      \global\StackPointer=\Iteration
      \Iteration=\LengthOfStack 
     \fi
    \fi
   \fi
  \advance\Iteration by 1
 \repeat}
}


\def\FindNext{%
 \global\StackPointer=-1 
 {\Iteration=0
  \loop
  \ifnum\Iteration<\LengthOfStack
   \GetItemSTATUS{\Iteration}
   \ifnum\ItemSTATUS=\InStack
    \GetItemTYPE{\Iteration}
   \ifnum\ItemTYPE=\Figure
    \ifMoreFigures
      \global\NextItem=\Figure
      \global\StackPointer=\Iteration
      \Iteration=\LengthOfStack 
    \fi
   \fi
   \ifnum\ItemTYPE=\Table
    \ifMoreTables
      \global\NextItem=\Table
      \global\StackPointer=\Iteration
      \Iteration=\LengthOfStack 
    \fi
   \fi
  \fi
  \advance\Iteration by 1
 \repeat}
}


\def\ChangeStatus#1#2{%
 \Point=\StatusStack
 \advance\Point by #1
 \global\count\Point=#2
}



\def\Zone{\InZoneA}

\ZoneBAdjust=0pt

\def\MakePage{
 \global\ZoneBSize=\PageHeight
 \global\TextSize=\ZoneBSize
 \global\ZoneAFullPagefalse
 \global\topskip=\TextLeading
 \MakePageInCompletetrue
 \MoreFigurestrue
 \MoreTablestrue
 \FigInZoneBfalse
 \FigInZoneCfalse
 \TabInZoneBfalse
 \TabInZoneCfalse
 \global\FirstSingleItemtrue
 \global\FirstZoneAtrue
 \global\setbox\ZoneABOX=\box\VOIDBOX
 \global\setbox\ZoneBBOX=\box\VOIDBOX
 \global\setbox\ZoneCBOX=\box\VOIDBOX
 \loop
  \ifMakePageInComplete
 \FindNext
 \ifnum\StackPointer=-1
  \NextItem=-1
  \MoreFiguresfalse
  \MoreTablesfalse
 \fi
 \ifnum\NextItem=\Figure
   \FindItem{\Figure}{\NextFigure}
   \ifnum\StackPointer=-1 \global\MoreFiguresfalse
   \else
    \GetItemSPAN{\StackPointer}
    \ifnum\ItemSPAN=\Single \def\Zone{\InZoneB}\relax
     \ifFigInZoneC \global\MoreFiguresfalse\fi
    \else
     \def\Zone{\InZoneA}
     \ifFigInZoneB \def\Zone{\InZoneC}\fi
    \fi
   \fi
   \ifMoreFigures\Print{}\FigureItems\fi
 \fi
\ifnum\NextItem=\Table
   \FindItem{\Table}{\NextTable}
   \ifnum\StackPointer=-1 \global\MoreTablesfalse
   \else
    \GetItemSPAN{\StackPointer}
    \ifnum\ItemSPAN=\Single\relax
     \ifTabInZoneC \global\MoreTablesfalse\fi
    \else
     \def\Zone{\InZoneA}
     \ifTabInZoneB \def\Zone{\InZoneC}\fi
    \fi
   \fi
   \ifMoreTables\Print{}\TableItems\fi
 \fi
   \MakePageInCompletefalse 
   \ifMoreFigures\MakePageInCompletetrue\fi
   \ifMoreTables\MakePageInCompletetrue\fi
 \repeat
 \ifZoneAFullPage
  \global\TextSize=0pt
  \global\ZoneBSize=0pt
  \global\vsize=0pt\relax
  \global\topskip=0pt\relax
  \vbox to 0pt{\vss}
  \eject
 \else
 \global\advance\ZoneBSize by -\ZoneBAdjust
 \global\vsize=\ZoneBSize
 \global\hsize=\ColumnWidth
 \global\ZoneBAdjust=0pt
 \ifdim\TextSize<23pt
 \Warn{}
 \Warn{* Making column fall short: TextSize=\the\TextSize *}
 \vskip-\lastskip\eject\fi
 \fi
}

\def\MakeRightCol{
 \global\TextSize=\ZoneBSize
 \MakePageInCompletetrue
 \MoreFigurestrue
 \MoreTablestrue
 \global\FirstSingleItemtrue
 \global\setbox\ZoneBBOX=\box\VOIDBOX
 \def\Zone{\InZoneB}
 \loop
  \ifMakePageInComplete
 \FindNext
 \ifnum\StackPointer=-1
  \NextItem=-1
  \MoreFiguresfalse
  \MoreTablesfalse
 \fi
 \ifnum\NextItem=\Figure
   \FindItem{\Figure}{\NextFigure}
   \ifnum\StackPointer=-1 \MoreFiguresfalse
   \else
    \GetItemSPAN{\StackPointer}
    \ifnum\ItemSPAN=\Double\relax
     \MoreFiguresfalse\fi
   \fi
   \ifMoreFigures\Print{}\FigureItems\fi
 \fi
 \ifnum\NextItem=\Table
   \FindItem{\Table}{\NextTable}
   \ifnum\StackPointer=-1 \MoreTablesfalse
   \else
    \GetItemSPAN{\StackPointer}
    \ifnum\ItemSPAN=\Double\relax
     \MoreTablesfalse\fi
   \fi
   \ifMoreTables\Print{}\TableItems\fi
 \fi
   \MakePageInCompletefalse 
   \ifMoreFigures\MakePageInCompletetrue\fi
   \ifMoreTables\MakePageInCompletetrue\fi
 \repeat
 \ifZoneAFullPage
  \global\TextSize=0pt
  \global\ZoneBSize=0pt
  \global\vsize=0pt\relax
  \global\topskip=0pt\relax
  \vbox to 0pt{\vss}
  \eject
 \else
 \global\vsize=\ZoneBSize
 \global\hsize=\ColumnWidth
 \ifdim\TextSize<23pt
 \Warn{}
 \Warn{* Making column fall short: TextSize=\the\TextSize *}
 \vskip-\lastskip\eject\fi
\fi
}

\def\FigureItems{
 \Print{Considering...}
 \ShowItem{\StackPointer}
 \GetItemBOX{\StackPointer} 
 \GetItemSPAN{\StackPointer}
  \CheckFitInZone 
  \ifnum\ItemFits=\Yes
   \ifnum\ItemSPAN=\Single
     \ChangeStatus{\StackPointer}{\InZoneB} 
     \global\FigInZoneBtrue
     \ifFirstSingleItem
      \hbox{}\vskip-\BodgeHeight
     \global\advance\ItemSIZE by \TextLeading
     \fi
     \unvbox\ItemBOX\ItemSep
     \global\FirstSingleItemfalse
     \global\advance\TextSize by -\ItemSIZE
     \global\advance\TextSize by -\TextLeading
   \else
    \ifFirstZoneA
     \global\advance\ItemSIZE by \TextLeading
     \global\FirstZoneAfalse\fi
    \global\advance\TextSize by -\ItemSIZE
    \global\advance\TextSize by -\TextLeading
    \global\advance\ZoneBSize by -\ItemSIZE
    \global\advance\ZoneBSize by -\TextLeading
    \ifFigInZoneB\relax
     \else
     \ifdim\TextSize<3\TextLeading
     \global\ZoneAFullPagetrue
     \fi
    \fi
    \ChangeStatus{\StackPointer}{\Zone}
    \ifnum\Zone=\InZoneC \global\FigInZoneCtrue\fi
  \fi
   \Print{TextSize=\the\TextSize}
   \Print{ZoneBSize=\the\ZoneBSize}
  \global\advance\NextFigure by 1
   \Print{This figure has been placed.}
  \else
   \Print{No space available for this figure...holding over.}
   \Print{}
   \global\MoreFiguresfalse
  \fi
}

\def\TableItems{
 \Print{Considering...}
 \ShowItem{\StackPointer}
 \GetItemBOX{\StackPointer} 
 \GetItemSPAN{\StackPointer}
  \CheckFitInZone 
  \ifnum\ItemFits=\Yes
   \ifnum\ItemSPAN=\Single
    \ChangeStatus{\StackPointer}{\InZoneB}
     \global\TabInZoneBtrue
     \ifFirstSingleItem
      \hbox{}\vskip-\BodgeHeight
     \global\advance\ItemSIZE by \TextLeading
     \fi
     \unvbox\ItemBOX\ItemSep
     \global\FirstSingleItemfalse
     \global\advance\TextSize by -\ItemSIZE
     \global\advance\TextSize by -\TextLeading
   \else
    \ifFirstZoneA
    \global\advance\ItemSIZE by \TextLeading
    \global\FirstZoneAfalse\fi
    \global\advance\TextSize by -\ItemSIZE
    \global\advance\TextSize by -\TextLeading
    \global\advance\ZoneBSize by -\ItemSIZE
    \global\advance\ZoneBSize by -\TextLeading
    \ifFigInZoneB\relax
     \else
     \ifdim\TextSize<3\TextLeading
     \global\ZoneAFullPagetrue
     \fi
    \fi
    \ChangeStatus{\StackPointer}{\Zone}
    \ifnum\Zone=\InZoneC \global\TabInZoneCtrue\fi
   \fi
  \global\advance\NextTable by 1
   \Print{This table has been placed.}
  \else
  \Print{No space available for this table...holding over.}
   \Print{}
   \global\MoreTablesfalse
  \fi
}


\def\CheckFitInZone{%
{\advance\TextSize by -\ItemSIZE
 \advance\TextSize by -\TextLeading
 \ifFirstSingleItem
  \advance\TextSize by \TextLeading
 \fi
 \ifnum\Zone=\InZoneA\relax
  \else \advance\TextSize by -\ZoneBAdjust
 \fi
 \ifdim\TextSize<3\TextLeading \global\ItemFits=\No
 \else \global\ItemFits=\Yes\fi}
}

\def\BF#1#2{
 \ItemSTATUS=\InStack
 \ItemNUMBER=#1
 \ItemTYPE=\Figure
 \if#2S \ItemSPAN=\Single
  \else \ItemSPAN=\Double
 \fi
 \setbox\ItemBOX=\vbox{}
}

\def\BT#1#2{
 \ItemSTATUS=\InStack
 \ItemNUMBER=#1
 \ItemTYPE=\Table
 \if#2S \ItemSPAN=\Single
  \else \ItemSPAN=\Double
 \fi
 \setbox\ItemBOX=\vbox{}
}

\def\BeginOpening{%
 \hsize=\PageWidth
 \global\setbox\ItemBOX=\vbox\bgroup
}

\let\begintopmatter=\BeginOpening  

\def\EndOpening{%
 \egroup
 \ItemNUMBER=0
 \ItemTYPE=\Figure
 \ItemSPAN=\Double
 \ItemSTATUS=\InStack
 \JoinStack
}


\newbox\tmpbox

\def\FC#1#2#3#4{%
  \ItemSTATUS=\InStack
  \ItemNUMBER=#1
  \ItemTYPE=\Figure
  \if#2S
    \ItemSPAN=\Single \TEMPDIMEN=\ColumnWidth
  \else
    \ItemSPAN=\Double \TEMPDIMEN=\PageWidth
  \fi
  {\hsize=\TEMPDIMEN
   \global\setbox\ItemBOX=\vbox{%
     \ifFigureBoxes
       \B{\TEMPDIMEN}{#3}
     \else
       \vbox to #3{\vfil}%
     \fi%
     \eightpoint\rm\bls{\rTenPT}%
     \vskip 5.5pt plus 6pt%
     \setbox\tmpbox=\vbox{#4\par}%
     \ifdim\ht\tmpbox>10pt 
       \noindent #4\par%
     \else
       \hbox to \hsize{\hfil #4\hfil}%
     \fi%
   }%
  }%
  \JoinStack%
  \Print{Processing source for figure {\the\ItemNUMBER}}%
}

\let\figure=\FC  

\def\TH#1#2#3#4{%
 \ItemSTATUS=\InStack
 \ItemNUMBER=#1
 \ItemTYPE=\Table
 \if#2S \ItemSPAN=\Single \TEMPDIMEN=\ColumnWidth
  \else \ItemSPAN=\Double \TEMPDIMEN=\PageWidth
 \fi
{\hsize=\TEMPDIMEN
\eightpoint\bls{\rTenPT}\rm
\global\setbox\ItemBOX=\vbox{\noindent#3\vskip 5.5pt plus5.5pt\noindent#4}}
 \JoinStack
 \Print{Processing source for table {\the\ItemNUMBER}}
}

\let\table=\TH  

\def\UnloadZoneA{%
\FirstZoneAtrue
 \Iteration=0
  \loop
   \ifnum\Iteration<\LengthOfStack
    \GetItemSTATUS{\Iteration}
    \ifnum\ItemSTATUS=\InZoneA
     \GetItemBOX{\Iteration}
     \ifFirstZoneA \vbox to \BodgeHeight{\vfil}%
     \FirstZoneAfalse\fi
     \unvbox\ItemBOX\ItemSep
     \LeaveStack{\Iteration}
     \else
     \advance\Iteration by 1
   \fi
 \repeat
}

\def\UnloadZoneC{%
\Iteration=0
  \loop
   \ifnum\Iteration<\LengthOfStack
    \GetItemSTATUS{\Iteration}
    \ifnum\ItemSTATUS=\InZoneC
     \GetItemBOX{\Iteration}
     \ItemSep\unvbox\ItemBOX
     \LeaveStack{\Iteration}
     \else
     \advance\Iteration by 1
   \fi
 \repeat
}


\def\ShowItem#1{
  {\GetItemAll{#1}
  \Print{\the#1:
  {TYPE=\ifnum\ItemTYPE=\Figure Figure\else Table\fi}
  {NUMBER=\the\ItemNUMBER}
  {SPAN=\ifnum\ItemSPAN=\Single Single\else Double\fi}
  {SIZE=\the\ItemSIZE}}}
}

\def\ShowStack{%
 \Print{}
 \Print{LengthOfStack = \the\LengthOfStack}
 \ifnum\LengthOfStack=0 \Print{Stack is empty}\fi
 \Iteration=0
 \loop
 \ifnum\Iteration<\LengthOfStack
  \ShowItem{\Iteration}
  \advance\Iteration by 1
 \repeat
}

\def\B#1#2{%
\hbox{\vrule\kern-0.4pt\vbox to #2{%
\hrule width #1\vfill\hrule}\kern-0.4pt\vrule}
}

\def\Ref#1{\begingroup\global\setbox\TEMPBOX=\vbox{\hsize=2in\noindent#1}\endgroup
\ht1=0pt\dp1=0pt\wd1=0pt\vadjust{\vtop to 0pt{\advance
\hsize0.5pc\kern-10pt\moveright\hsize\box\TEMPBOX\vss}}}

\def\MarkRef#1{\leavevmode\thinspace\hbox{\vrule\vtop
{\vbox{\hrule\kern1pt\hbox{\vphantom{\rm/}\thinspace{\rm#1}%
\thinspace}}\kern1pt\hrule}\vrule}\thinspace}%


\output{%
 \ifLeftCOL
  \global\setbox\LeftBOX=\vbox to \ZoneBSize{\box255\unvbox\ZoneBBOX}
  \global\LeftCOLfalse
  \MakeRightCol
 \else
  \setbox\RightBOX=\vbox to \ZoneBSize{\box255\unvbox\ZoneBBOX}
  \setbox\MidBOX=\hbox{\box\LeftBOX\hskip\ColumnGap\box\RightBOX}
  \setbox\PageBOX=\vbox to \PageHeight{%
  \UnloadZoneA\box\MidBOX\UnloadZoneC}
  \shipout\vbox{\PageHead\box\PageBOX\PageFoot}
  \global\advance\pageno by 1
  \global\HeaderNumber=\DefaultHeader
  \global\LeftCOLtrue
  \CleanStack
  \MakePage
 \fi
}


\catcode `\@=12 
